\documentclass[twocolumn,twoside]{revtex4}
\usepackage{hyperref}
\usepackage{graphicx}
\usepackage{fancyhdr}
\usepackage[utf8]{inputenc}
\usepackage{lipsum}
\usepackage{lineno}
\begin{document}
\title{Searching for leptoquarks with the ATLAS detector}

%

\author{Vincent Wai Sum Wong\\
on behalf of the ATLAS Collaboration}
\affiliation{Department of Physics and Astronomy, University of British Columbia, BC, Canada, V6T1Z1}

\begin{abstract}
Results from the latest searches for pair-produced scalar leptoquarks using 36.1 $\text{fb}^{-1}$ of $pp$-collision data recorded by the ATLAS detector at $\sqrt{s}$ = 13 TeV were presented. No statistically significant excess of data over Standard Model prediction is observed. The observed limits on first- (second-) generation leptoquark masses are excluded up to 1400 (1560) GeV in the minimal Buchm\"uller-R\"uckl-Wyler model, assuming a leptoquark decay branching ratio of 100\% into a charged lepton and a quark. Third generation leptoquark masses are excluded up to 1000 GeV at the highest and lowest decay branching ratios for both up-type and down-type leptoquarks.
\end{abstract}

\maketitle

\thispagestyle{fancy}

\section{Introduction}
The quark and lepton sectors of the Standard Model (SM) are interestingly similar, motivating one to hypothesize a fundamental symmetry between the two sectors. Such a symmetry can be found in many grand unified theories, such as grand unified SU(5), Pati-Salam models based on SU(4) or R-parity violating (RPV) supersymmetry (SUSY) models. These models naturally predict a new class of bosons carrying both lepton and baryon number, called leptoquarks (LQs). LQs are hypothetical colour triplet bosons, which couple directly to a quark and a lepton. They can be of either scalar or vector nature and they carry fractional electromagnetic charge. Recently, LQs have gained more attention as they could provide an attractive explanation to the recent hint of lepton flavour universality violation from the observed B decay anomalies in BaBar~\cite{babar}, Belle~\cite{belle} and LHCb~\cite{lhcb1}~\cite{lhcb2}~\cite{lhcb3}. If their mass were near the TeV scale, leptoquarks could be produced at the Large Hadron Collider~\cite{lhc}. 

The ATLAS collaboration has performed searches for first-, second- and third-generation LQ pair production using 36.1 $\text{fb}^{-1}$ of proton-proton collision data recorded at a centre-of-mass energy $\sqrt{s}$ = 13 TeV~\cite{first_second_lq}~\cite{third_lq}.\makeatletter{\renewcommand*{\@makefnmark}{}
\footnotetext{\textcopyright  \ Copyright 2019 CERN for the benefit of the ATLAS Collaboration CC-BY-4.0 license.}\makeatother}The searches use the minimal Buchm\"uller-R\"uckl-Wyler (mBRW) model~\cite{mBRW} as a benchmark, in which LQs are assumed to interact with leptons and quarks of the same generation only. LQs couple to the lepton-quark pair via a Yukawa coupling, and the coupling is modelled by two parameters: (i) coupling constant $\lambda$ and (ii) the branching fraction of LQ decays involving charged lepton $\beta$. The coupling to a charged lepton and a quark is given by $\sqrt{\beta}\lambda$, while the coupling to a neutrino and a quark is $\sqrt{1-\beta}\lambda$. LQ pair production is largely driven by the strong interaction and thus independent of $\lambda$. All analyses presented here assume that LQs are scalar, so LQ mass and $\beta$ remain the two important free parameters in the model of LQ pair production. More details about the Monte Carlo (MC) simulation samples used for signal and background modelling can be found in Refs.~\cite{first_second_lq,third_lq}. 

\section{The ATLAS detector}
The ATLAS detector~\cite{atlas} is a general-purpose cylindrical detector, which has nearly 4$\pi$ solid angle coverage. Closest to the beam line is the inner tracking detector surrounded by a solenoid magnet with a homogeneous field of 2 T. Next are the electromagnetic and hadronic calorimeters that are used for particle shower reconstruction and energy measurement. Outside the calorimeters is the muon spectrometer with a toroidal magnet. All of these subsystems are used in the LQ searches. 

Electrons are reconstructed using clusters of electromagnetic-calorimeter cells with significant energy deposits matched to tracks that are reconstructed in the inner detector. Muons are reconstructed by a combined measurement of tracks in the inner detector and the muon spectrometer. Hadronic jets are reconstructed from clusters of energy deposits in the calorimeters using the anti-$k_{t}$ clustering algorithm~\cite{antikt} with radius parameter of 0.4. To identify jets originating from $b$-hadrons, a multivariate technique is employed to recognize the presence of secondary vertices sufficiently displaced with respect to the primary interaction vertex. Hadronically decaying $\tau$-leptons are identified from the inner detector tracks and the calorimeter shower shapes using a multivariate algorithm. The missing transverse energy is calculated as the negative of the vector sum of all calibrated hard objects plus soft contributions from preselected tracks associated with the most energetic vertex of the event. 

\section{Search for first- and second-generation LQs}
A search for the pair production of first- and second-generation LQs was performed, with focus on LQs decaying to a pair of charged leptons and jets ($\ell \ell jj$), or one charged lepton, a neutrino and jets ($\ell \nu jj$). Electron and muon channels are treated individually to search for first- and second-generation LQs respectively, but they share a very similar analysis strategy. In both of the electron and muon channels, $\ell \ell jj$ and $\ell \nu jj$ channels are combined to improve sensitivity to various values of $\beta$. The candidate events in the $\ell \ell jj$ channel can be characterized by exactly two same-flavour leptons and at least two jets, while the $\ell \nu jj$ channel requires exactly one lepton, missing transverse momentum and at least two jets. 

\begin{figure*}[!htpb]
\centering
\includegraphics[width=70mm]{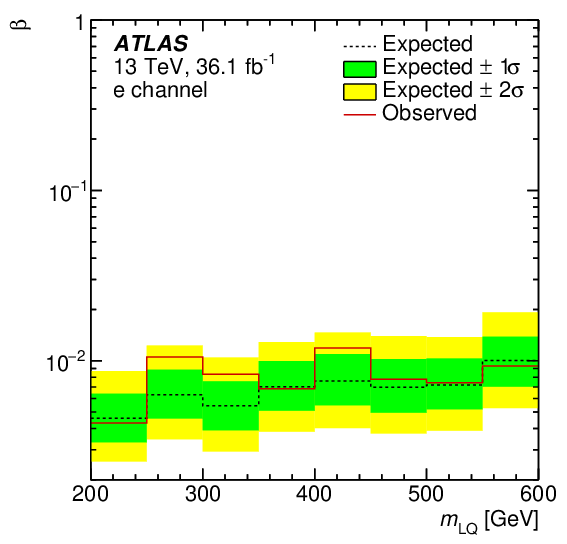}
\includegraphics[width=70mm]{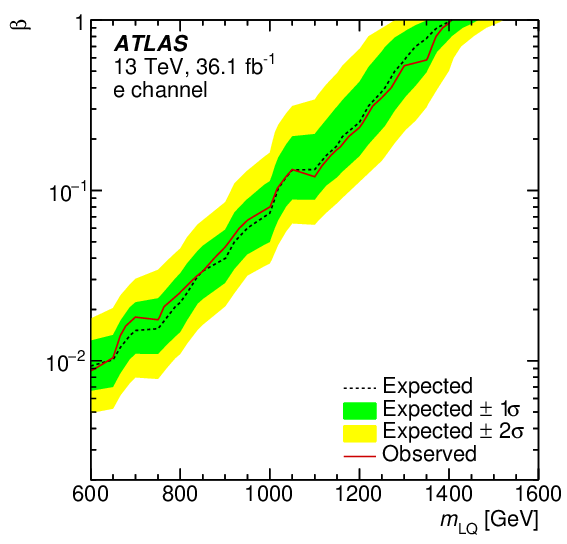}
\caption{Upper exclusion limits at 95\% CL on the signal cross-section for pair production of first-generation LQs in the $\text{m}_{\text{LQ}}$-$\beta$ plane ~\cite{first_second_lq}. The solid red (dashed black) curve shows the observed (expected) exclusion limit.} \label{fig:lq1Limit}
\end{figure*}

\begin{figure*}[!htpb]
\centering
\includegraphics[width=70mm]{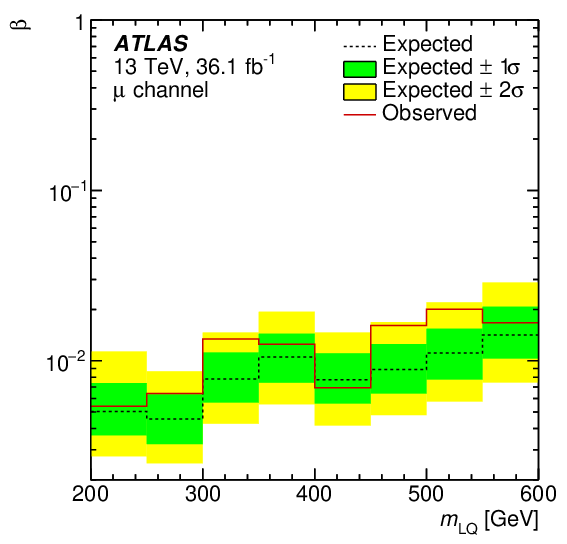}
\includegraphics[width=70mm]{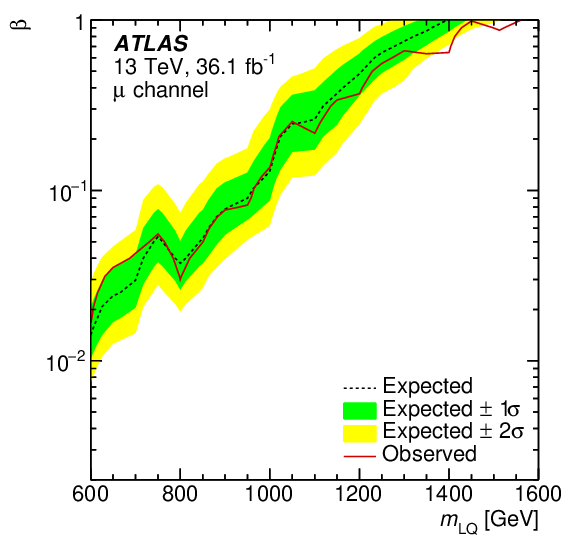}
\caption{Upper exclusion limits at 95\% CL on the signal cross-section for pair production of second-generation LQs in the $\text{m}_{\text{LQ}}$-$\beta$ plane ~\cite{first_second_lq}. The solid red (dashed black) curve shows the observed (expected) exclusion limit.} \label{fig:lq2Limit}
\end{figure*}

The dominant background in the $\ell \ell jj$ channel is from $Z/\gamma*$+jets, which is suppressed by applying high lepton invariant mass cuts $m_{\ell \ell} >$ 130 GeV. In the $\ell \nu jj$ channel, $t\bar{t}$ and $W$+jets are two of the major backgrounds. To reduce these backgrounds, the transverse mass of lepton and missing transverse energy is required to be above 130 GeV. There are also reducible backgrounds due to objects being misidentified as leptons or non-prompt leptons that are produced from hadron decays. These are collectively called fake (lepton) background and are estimated with a data-driven approach. A missing transverse energy-related quantity, defined by $S = E_{T}^{\text{miss}}/\sqrt{p_{T}^{j1}+p_{T}^{j2}+p_{T}^{\ell}}$, is used to suppress the contributions from fake background. In order to obtain a better background estimation, a reweighting of the $Z/\gamma*$+jets and $W$+jets simulation is performed. The weights in the reweighting procedure are parameterised as a function of dijet mass in dedicated control regions with enriched $Z/\gamma*$+jets and $W$+jets events respectively. To describe the data better, these control regions and an additional control region enriched in $t\bar{t}$ events are used to constrain the normalizations of these main background sources simultaneously. Finally, boosted decision trees (BDTs) are used to identify signal from background after training on discriminating variables, for instance the LQ candidate masses and the transverse momentum of jet and lepton. In the $\ell \ell jj$ channel, LQ candidates are reconstructed by minimizing the difference between the masses of the two lepton-jet pairs. In the $\ell \nu jj$ channel, LQ candidates are reconstructed by minimizing the difference between the masses of the lepton-jet pair and the missing-energy-jet-pair.

In each of the four channels, an optimized cut is applied on the BDT score to define the corresponding signal region, and the amount of signal is extracted from data simultaneously in the $\ell \ell jj$ and $\ell \nu jj$ channel. Good agreement is observed between data and background predictions in all channels. Limits are set at 95\% confidence level on the cross-section times the square of branching ratio to charged lepton and quark. Exclusion contours in the $\text{m}_{\text{LQ}}$-$\beta$ plane for electron and muon channel are shown in Figure~\ref{fig:lq1Limit} and~\ref{fig:lq2Limit} respectively. The observed lower limit on the masses of first- and second-generation LQs is 1.4 TeV.

\section{Search for third-generation LQs}
In the search for third-generation LQs, two types of LQs are defined: up-type and down-type. Up-type LQs decay to a top quark and a neutrino, or a bottom quark and a tau. Down-type LQs decay to a bottom quark and a neutrino, or a top quark and a tau. The search presented here is based on the reinterpretations of four ATLAS searches for SUSY particles, and a dedicated search for LQ$\rightarrow b\tau$. 

SUSY particles can give identical or very similar signatures to the ones from LQ pair production. In the R-parity-conserving SUSY framework, the top (bottom) squark is produced in pairs and it decays into a SM top (bottom) quark and the lightest SUSY particle which is weakly interacting. Hence, the ATLAS searches for top squarks in events with 0 or 1 lepton~\cite{stop_0lep}~\cite{stop_1lep} and searches for bottom squarks~\cite{sbottom} are optimal for up-type and down-type LQs with $\beta$=0. Another top squark pair production search is also re-interpreted in which the top squarks decay to tau sleptons~\cite{stop2stau}. The final state of such a scenario contains two $\tau$-leptons, $b$-jets and missing transverse momentum. The reinterpretation of this SUSY search for LQs is expected to be sensitive to all $\beta$ values except near zero, with the best sensitivity at medium $\beta$ value. 

The dedicated analysis searching for $b\tau b\tau$ final state targets the fully hadronic and semi-leptonic $\tau$-lepton decays. Due to the similarity to the search for di-Higgs decaying into $bb\tau \tau$~\cite{dihiggs}, the same strategy is used with re-optimization for LQs. The candidate events with fully hadronic tau decays can be characterized by exactly two $\tau$-leptons and at least two jets, without any electron or muon, whereas the candidate events with semi-leptonic tau decays require exactly one $\tau$-lepton, one electron or muon and at least two jets. In both fully hadronic and semi-leptonic channels, the candidate events are categorized into 1 $b$-tagged and 2 $b$-tagged regions. Similar to the first- and second-generation LQs search, the LQ candidates are constructed by minimizing the mass difference between the two $\tau$-jet pairs (the $\tau$-jet pair and the $\ell$-jet pair) in the $\tau_{\text{had}}\tau_{\text{had}}$ ($\tau_{\text{lep}}\tau_{\text{had}}$) channel. The dominant backgrounds in this search are  $t\bar{t}$ and $Z/\gamma*(\rightarrow \tau \tau)$+heavy-flavour jets events, as well as multi-jet events with misidentified $\tau$'s estimated from data. BDTs are trained on a set of observables, such as the LQ candidate masses and the angular separation of $\tau$-jet pair or lepton-jet pair, to classify signal from background. To optimize the sensitivity, a separate BDT is trained for each of the four different signal regions and for each LQ mass. Finally, a single combined fit of the BDT output score profiles in all signal and control regions is used to extract the evidence of signal events. 

\begin{figure*}[!htpb]
\centering
\includegraphics[width=80mm]{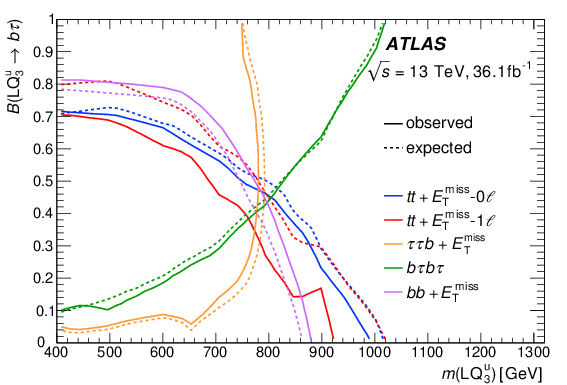}
\includegraphics[width=80mm]{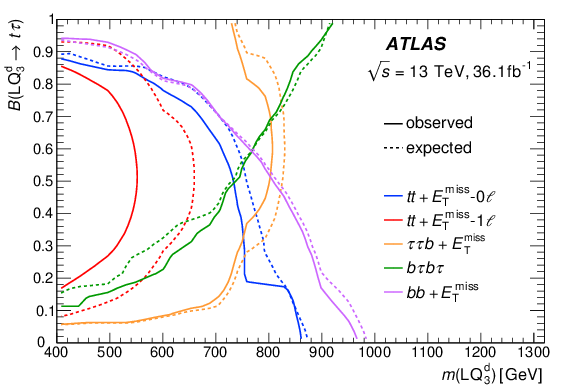}
\caption{Upper exclusion limits at 95\% CL on the signal cross-section for third-generation up-type (left) and down-type (right) leptoquark pair production in the $\text{m}_{\text{LQ}}$-branching-ratio plane ~\cite{third_lq}. The solid (dashed) curves shows the observed (expected) exclusion limits, which are based on four reinterpretations of SUSY searches and a dedicated LQ search for final state of two $b$-jets and two $\tau$-leptons.} \label{fig:lq3Limit}
\end{figure*}

No significant excess above SM expectation is observed, and limits are set on the LQ mass as a function of branching ratio for all five signatures as shown in Figure~\ref{fig:lq3Limit}. The observed lower limit on the masses of third-generation LQs is 800 GeV for both up-type and down-type LQs, regardless of the branching ratio. At the extrema of branching ratio equal to zero or one, masses below around 1000 GeV are excluded.

\section{Conclusion}
The ATLAS collaboration has performed searches for first-, second- and third-generation scalar LQs. No significant excess above SM expectation has been observed with 36.1 $\text{fb}^{-1}$ of $pp$ collision data taken in 2015 and 2016 at $\sqrt{s}$=13 TeV. Exclusion limits on LQ masses as a function of branching ratio have been interpreted as results in the benchmark LQ model.

\bigskip 
\newpage
\bibliography{fpcp2019_vwong}
\bibliographystyle{ieeetr}



\end{document}